\documentclass{aa}
\usepackage{graphicx}
\usepackage{txfonts}

\usepackage{lineno}

\begin{document}

   \title{Looking for blazars in a sample of unidentified high-energy
   emitting Fermi sources}

   \author{E. J. Marchesini \inst{1,2,3,4,5} \and N. Masetti \inst{5,6} \and V. Chavushyan \inst{7} \and S. A. Cellone \inst{1,2} \and I. Andruchow \inst{1,2} \and L. Bassani \inst{5} \and A. Bazzano \inst{8} \and E. Jim\'enez-Bail\'on \inst{9} \and R. Landi \inst{5} \and A. Malizia \inst{5} \and E. Palazzi \inst{5} \and V. Pati\~no-\'Alvarez \inst{7} \and G. A. Rodr\'iguez-Castillo \inst{10} \and J. B. Stephen \inst{5} \and P. Ubertini \inst{8}}
   
   \institute{Facultad de Ciencias Astron\'omicas y Geof\'isicas, Universidad Nacional de La Plata, Paseo del Bosque, B1900FWA, La Plata, Argentina. \and
   Instituto de Astrof\'isica de La Plata, CONICET--UNLP, CCT La Plata, Paseo del Bosque, B1900FWA, La Plata, Argentina. \and
   Dipartimento di Fisica, Universit\`a degli Studi di Torino, via Pietro Giuria 1, I-10125 Torino, Italia. \and
   INFN -- Istituto Nazionale di Fisica Nucleare, Sezione di Torino, via Pietro Giuria 1, I-10125 Torino, Italia.\and
   INAF -- Istituto di Astrofisica Spaziale e Fisica Cosmica di Bologna, via Gobetti 101, I-40129, Bologna, Italia. \and
   Departamento de Ciencias F\'isicas, Universidad Andr\'es Bello, Fern\'andez Concha 700, Las Condes, Santiago, Chile. \and
   Instituto Nacional de Astrof\'isica, \'Optica y Electr\'onica, Apartado Postal 51-216, 72000 Puebla, M\'exico. \and
   INAF--Istituto di Astrofisica e Planetologia Spaziali, Via Fosso del Cavaliere 100, I-00133-Rome, Italia. \and
   Instituto de Astronom\'ia, Universidad Nacional Aut\'onoma de M\'exico, Apdo. Postal 877, Ensenada, 22800 Baja California, M\'exico. \and
   Osservatorio Astronomico di Roma, INAF, via Frascati 33, I-00040 Monteporzio Catone, Italia.}
   \date{}

%
  \abstract
   {Based on their overwhelming dominance among associated Fermi $\gamma$--ray catalogue sources, it is expected that a large fraction of the unidentified Fermi objects are blazars. Through crossmatching
   between the positions of unidentified $\gamma$--ray sources from the First Fermi
   Catalog of $\gamma$--ray sources emitting above 10 GeV (1FHL) and the
   ROSAT and Swift/XRT catalogues of X--ray objects and between pointed XRT observations, a sample
   of 36 potential associations was found in previous works with less than 15 arcsec of positional offset. One-third of them have recently been classified; the remainder,
   though believed to belong to the blazar class, still lack spectroscopic classifications.}
   {We study the optical spectrum of the putative counterparts of these
   unidentified gamma-ray sources in order to find their redshifts and to determine
   their nature and main spectral characteristics.}
   {An observational campaign was carried out on the putative counterparts of 13 1FHL sources using medium--resolution optical spectroscopy from the Osservatorio Astronomico di Bologna in Loiano, Italy; the Telescopio
   Nazionale Galileo and the Nordic Optical Telescope, both in the Canary Islands, Spain; and the Observatorio Astron\'omico Nacional San Pedro M\'artir in Baja California, Mexico.}
   {We were able to classify 14 new objects based on their continuum shapes and
   spectral features.}
   {Twelve new blazars were found, along with one new quasar and one new
   narrow line Seyfert 1 (NLS1) to be potentially associated with the 1FHL sources of our sample. Redshifts or lower limits were obtained when
   possible alongside central black hole mass and luminosity estimates for the NLS1 and the quasar.
   }

   \maketitle
%
%
\section{Introduction}

The most important objective of the Fermi mission is to study the whole sky at $\gamma$--ray energies; this is achievable with the use of the Large Area Telescope (LAT) thanks to its large collecting area and field of view \citep{Atwood09}. The location accuracy of the telescope, which detects $\gamma$--ray objects
emitting at GeV energies, is between 0.5 to 10 arcminutes, depending on the source detection significance.

There are more than 3000 sources listed in the latest release of the Fermi catalogue \citep{Acero15}. Of these, only 238 are considered firm identifications by the LAT team, based on spatial morphology,
correlated variability, and/or periodic lightcurve properties. Another
$\sim 1800$ sources have high confidence associations, based on
cross-correlations with multiwavelength catalogues. The majority of these identified and associated sources belong to one of the following categories: extragalactic objects such as blazars (flat spectrum radio
quasars or BL Lacs), or Galactic sources (mainly pulsars, pulsar wind
nebulae, and supernova remnants). However, there is still an important number of sources (about 30\%) without proper identification, i.e. lacking association with any known class of $\gamma$--ray emitting objects, which constitute the class of unidentified/unassociated gamma-ray sources (UGSs).

A similar but less critical situation is found when considering the First Fermi Catalog of detected sources above 10 GeV \citep[1FHL;][]{Ackermann13}: from a total of 514 listed sources, 65 ($\sim$13\%) are UGSs. These are also the numbers resulting from analysing the Second Fermi Catalog of detected sources above 50 GeV \citep[2FHL;]{Ackermann16}: it lists 360 sources, of which 48 (14\%) are UGSs.

The search for counterparts of these new high--energy sources is hindered by the relatively large (in comparison with longer wavelengths) Fermi positional error ellipses. This uncertainty in their location means that positional correlations with known objects is often not enough to
identify a Fermi source; thus, a multiwavelength approach is needed in order to understand their nature, using X--ray, optical and radio data of likely counterparts. 
X--ray data analyses are particularly useful in finding a positionally correlated
object with broadband spectral parameters that might be
expected in a $\gamma$--ray emitting source. Soft X--ray surveys (i.e. with energies below 10 keV) are convenient for this task because
they offer 3 great advantages: They cover the whole
Fermi error ellipse, their positional accuracy is of the order of arcseconds, and
 they provide information in an energy band close to that at which the
Fermi LAT operates. Since most of the 1FHL sources are BL Lacs and in particular high-energy cutoff BL Lacs (HBL), and as they show the peak of the SED synchrotron component in the X-rays, crossmatching the Fermi catalogue with X-ray surveys should prove useful as a tool to select them. This allows the positional
uncertainty of the objects detected with Fermi to be restricted, thus facilitating the
identification process.

To this end, following \citet{Stephen10}, \citet{Landi15c,Landi15a,Landi15b} performed a crossmatch between the positions in the 1FHL
catalogue, the ROSAT All--Sky Survey Bright Source Catalogue of sources detected between 0.1--2.4 keV \citep{Voges99} , the 1SXPS Catalogue of X--ray sources detected with Swift/XRT in the 0.3-–10 keV band \citep{Evans14}, and pointed XRT observations available at the \rm{ASI} Science Data Center\footnote{http://www.asdc.asi.it/} archive. 
They found correlations with a strong level of confidence ($\sim 90\%$), leading to evidence for the potential association of a number of UGSs with X--ray counterparts, improving the positional error in all correlated objects, and thus opening the possibility for optical follow--up.

In particular, 36 secure 1FHL/X--ray potential associations were obtained which allowed the selection of a likely low--energy (optical and below) counterpart for all of
them. An investigation of the nature of these sources on the
basis of their archival multiwavelength properties indicates that all
potential associations are either recently identified blazars \citep{Landi15b,Landoni15,Massaro15,Ricci15} or blazar candidates \citep{Landi15a,Landi15c}. The majority of blazars are expected to show $\gamma$--ray emission in the GeV range \citep[e.g.][]{Acero15}. Nevertheless, 24 of the potential 36 associations are still
lacking an optical spectroscopic confirmation of their nature. 

According to \citet{Stephen10} and \citet{Landi15a}, 1FHL sources like these can be responsible for the emission of very
high energy $\gamma$--rays, up to the teraelectronvolt (TeV) range \citep{PadovaniGiommi95,Fossati98}. The interest in
extreme TeV blazars arises from the possibility of obtaining
information on both the acceleration processes of charged particles in
relativistic flows \citep[e.g.][]{Ghisellini10} and the intensity of
the extragalactic background light \citep[e.g.][]{Georganopoulos10}, which reflects the time--integrated
history of light production and re--processing in the universe, and hence
its measurement can provide information on the history of cosmological
star formation \citep{Mankuzhiyil10}. This is important when
considering that in the 1FHL catalogue, only 22 ($<\,6$\%) objects of the AGN type are considered to be firmly identified out of a total of 393 cases \citep{Ackermann13}. This is why the
confirmation of the nature of even a small subset of the unidentified
objects of the 1FHL sample would significantly increase the
statistics of the GeV/TeV emitting blazars class, which in turn is only achievable after finding the proper association. This would also be relevant for a future search of TeV blazars that can be performed with the Cherenkov Telescope Array \citep{Massaro13b}.

Furthermore, as the number of detected sources in the high--energy surveys is growing at an ever--increasing speed, it is necessary to establish well--defined methods to correctly identify and classify as many objects as possible while strictly reducing their positional uncertainties. Therefore, the aim of this work is to spectroscopically analyse 14 optical targets with near-positional coincidence with the X--ray sources out of those 24 without classification. Following the treatment of \citet{Stephen10}, we expect no more than only one spurious correlation out of the selected sample of 14 objects.

   \begin{table*}[!htp]
   
   \caption{Observed sample of unidentified sources from the 1FHL catalogue.} 
   \label{tab1} 
   \centering
      \setlength{\tabcolsep}{2pt}

   \begin{tabular}{cccccccc}
   \hline\hline
   
 Number & USNO designator &  \rm{RA(J2000)} & \rm{DEC(J2000)} & Observatory & UT date  & Time  & Total exp. \\ 
        & X-ray association & 				&				&				&[mm/dd/yy]	& [mid. exp.]	&  [s] \\
    (1) & (2) & (3) & (4) & (5) & (6) & (7) & (8) \\
   \hline\hline

1 & U0750-00173701		& $00^h43^m48\fs66$ & $-11^{\circ}16'07\farcs2$ & NOT & 10/13/2015 & $02:46:41$ & 1200 \\
  & 1RXS J004349.3-111612 & &&&&& \\
\hline
2 & U0975-00792795		& $03^h38^m29\fs24$ & $+13^{\circ}02'15\farcs2$ & NOT & 10/13/2015 & $04:59:08$ & 1200 \\
  & 1SXPS J033829.0+130213 & &&&&& \\
\hline
3 & U0675-01653184		& $04^h39^m49\fs54^s$ & $-19^{\circ}01'02\farcs5$ & NOT & 10/13/2015 & $05:35:13$ & 1200 \\
  & 1SXPS J043949.5-190102 & &&&&& \\
\hline
4 & U0750-02519189		& $06^h40^m07\fs31^s$ & $-12^{\circ}53'18\farcs6$ & NOT & 10/13/2015 & $06:08:22$ & 1200 \\
  & 1SXPS J064007.1-125313 & &&&&& \\
\hline
5 & U0875-0218538		& $07^h46^m27\fs14^s$ & $-02^{\circ}25'50\farcs7$ & NOT & 10/14/2015 & $04:53:16$ & 1200 \\
  & 1SXPS J074627.1-022550 & &&&&& \\
\hline
6 & U0825-05946383		& $08^h04^m57\fs74^s$ & $-06^{\circ}24'26\farcs3$ & LOI & 03/10/2015 & $21:21:16$ & 3600 \\
  & 1RXS J080458.3-062432 & &&&&& \\
\hline
7 & U0825-05946383		& $11^h15^m15\fs58^s$ & $-07^{\circ}01'25\farcs6$ & SPM & 01/14/2016 & $12:31:40$ & 1800 \\
  & SWXRT J111515.3-070126 & &&&&& \\
\hline
8 & U0750-08080787		& $13^h15^m52\fs98^s$ & $-07^{\circ}33'02\farcs0$ & LOI & 03/13/2015 & $00:57:54$ & 3600 \\
  & 1SXPS J131553.0-073301 & &&&&& \\
\hline
9 & U1575-03416792  	& $14^h10^m45\fs83^s$ & $+74^{\circ}05'11\farcs1$ & TNG & 08/26/2015 & $22:36:36$ & 2400 \\
  & 1SXPS J141045.3+740508 & &&&&& \\
\hline
10 & U1575-03416943  & $14^h10^m52\fs03^s$ & $+74^{\circ}04'15\farcs1$ & TNG & 08/26/2015 & $23:40:59$ & 1200 \\
   & 1SXPS J141051.3+740410 & &&&&& \\
\hline
11 & U0600-17715078	& $15^h12^m12\fs76^s$ & $-22^{\circ}55'08\farcs4$ & TNG & 08/27/2015 & $22:41:45$ & 2000 \\
   & 1RXS J151213.1-225515 & &&&&& \\
\hline
12 & U0825-08948904	& $15^h49^m52\fs17^s$ & $-06^{\circ}59'08\farcs3$ & TNG & 08/27/2015 & $23:30:37$ & 2400 \\ 
   & 1SXPS J154952.1-065908 & &&&&& \\
\hline
13 & U1125-10089754	& $18^h41^m21\fs72^s$ & $+29^{\circ}09'41\farcs2$ & TNG & 08/27/2015 & $00:24:02$ & 1600 \\
   & 1RXS J184121.8+290932 & &&&&& \\
\hline
14 & U1530-0317394	& $20^h02^m45\fs36^s$ & $+63^{\circ}02'33\farcs6$ & TNG & 08/29/2015 & $00:03:54$ & 3600 \\
   & 1RXS J200245.4+630226 & &&&&& \\
   \hline\hline

   \end{tabular}
   \caption{We report the name in the USNO and X--rays catalogues in column 2, in columns 3 and 4 coordinates referring to J2000.0 for each optical target, in column 5 the observatory, in column 6 the date of observation, in column 7 the UT time at mid exposure, and in column 8 the total exposure time in seconds for each of the optical pointings.}
   \end{table*}

In the following sections, we describe our optical follow--up work on a subsample of 14 of the aforementioned potentially associated objects from the 1FHL catalogue. From these, only 1FHL\,J1549.9-0658 appears in the 2FHL catalogue (named 2FHL\,J1549.8-0659), although there is also a detection positionally consistent (2FHL\,J0639.9-1252, at a distance of $\sim 3$ arcmin) with 1FHL\,J0639.6-1244. The reason why only one of the 1FHL objects from our sample can be found in the 2FHL catalogue is the energy threshold: The 2FHL catalogue includes only those sources detected at 50 GeV or more, while the 1FHL catalogue has a threshold of 10 GeV.

We note that 1FHL\,J1410.4+7408 shows two different X--ray objects \citep{Landi15a} within its $\gamma$--ray positional error box, each with a single corresponding optical source. We define 1FHL\,J1410.4+7408 A as the one marked as \#1 in \citet{Landi15a}, and 1FHL\,J1410.4+7408 B as the one marked as \#2. In section 2 we briefly discuss the selection of the sample, in section 3 we describe the observations, in section 4 we analyse our results, and in section 5 we summarise our conclusions.

\section{Sample selection}

Our sample of 1FHL fields is a subset of those presented in \citet{Landi15c,Landi15a}. 

They found only one X--ray counterpart for each Fermi source, with the exception of 1FHL\,J1410.4+7408. However, despite the better positional accuracy achieved, it is important to note that X--ray error circles are still large enough (i.e. $\sim$6 arcseconds) to find more than one optical source tentatively associated with each single X--ray counterpart. Thus, a supplementary investigation is needed to single out the actual counterpart of the $\gamma$--ray/X--ray emitter. For this reason, we set up an international campaign to obtain spectroscopic observations of candidate optical counterparts in 13 fields, which are the subject of this paper. Details on the observations can be found in Table 1.

\section{Observations}

The optical spectroscopic observations were carried out at four different
observatories for a total of 18 nights:

\begin{itemize} 
\item Three nights (from 10 Mar 2015 to 12 Mar 2015) at the 1.52m Cassini telescope of the Bologna Observatory in Loiano (LOI), Italy, with the BFOSC spectrograph and a 2.0 arcsec slit (0.40 nm/px dispersion). The data covered a range from 350 to 800\,nm.
\item Three nights (19 May 2015, 21 Jun 2015, and 09 Jul 2015) at the 3.58m Telescopio Nazionale Galileo (TNG) in La Palma, Canary Islands, Spain, with the DOLORES (LRS) spectrograph and a 1.5 arcsec slit (0.25 nm/px dispersion). The data covered a range from 370 to 800\,nm.
\item Two nights (13 Oct 2015 and 14 Oct 2015) at the 2.5m\,Nordic Optical Telescope (NOT), in La Palma, Canary Islands, Spain, with the ALFOSC spectrograph and a 1.0 arcsec slit (0.30 nm/px dispersion). The data
covered the 350 to 900\,nm range.
\item Eight nights (from 06 Nov 2015 to 09 Nov 2015 and from 14 Jan 2016 to
17 Jan 2016) at the 2.12m telescope in San Pedro M\'artir (SPM), Mexico, with the Boller \& Chivens spectrograph and a 2.5 arcsec slit (0.23 nm/px dispersion). The data covered a range from 350 to 800\,nm.
\end{itemize}

The data were cleaned from cosmic rays, bias corrected, flat--fielded, and both wavelength and flux calibrated using IRAF\footnote{IRAF is distributed by the National Optical Astronomy Observatory, which is operated by the Association of Universities for Research in Astronomy (AURA) under a cooperative agreement with the National Science Foundation.} standard packages, wavelength calibration lamps, and spectrophotometric standard stars. In each case, the estimated wavelength calibration error is less than 0.4\,nm.

\section{Results}

In Figure \ref{fig1}, we present the optical spectra for each analysed object in the upper panels, while in the lower panels we show the continuum--normalised spectra in order to highlight the presence of spectral features (if any). 


   \begin{figure*}[!htp] \centering
   
     \includegraphics[angle=270,width=0.463\hsize]{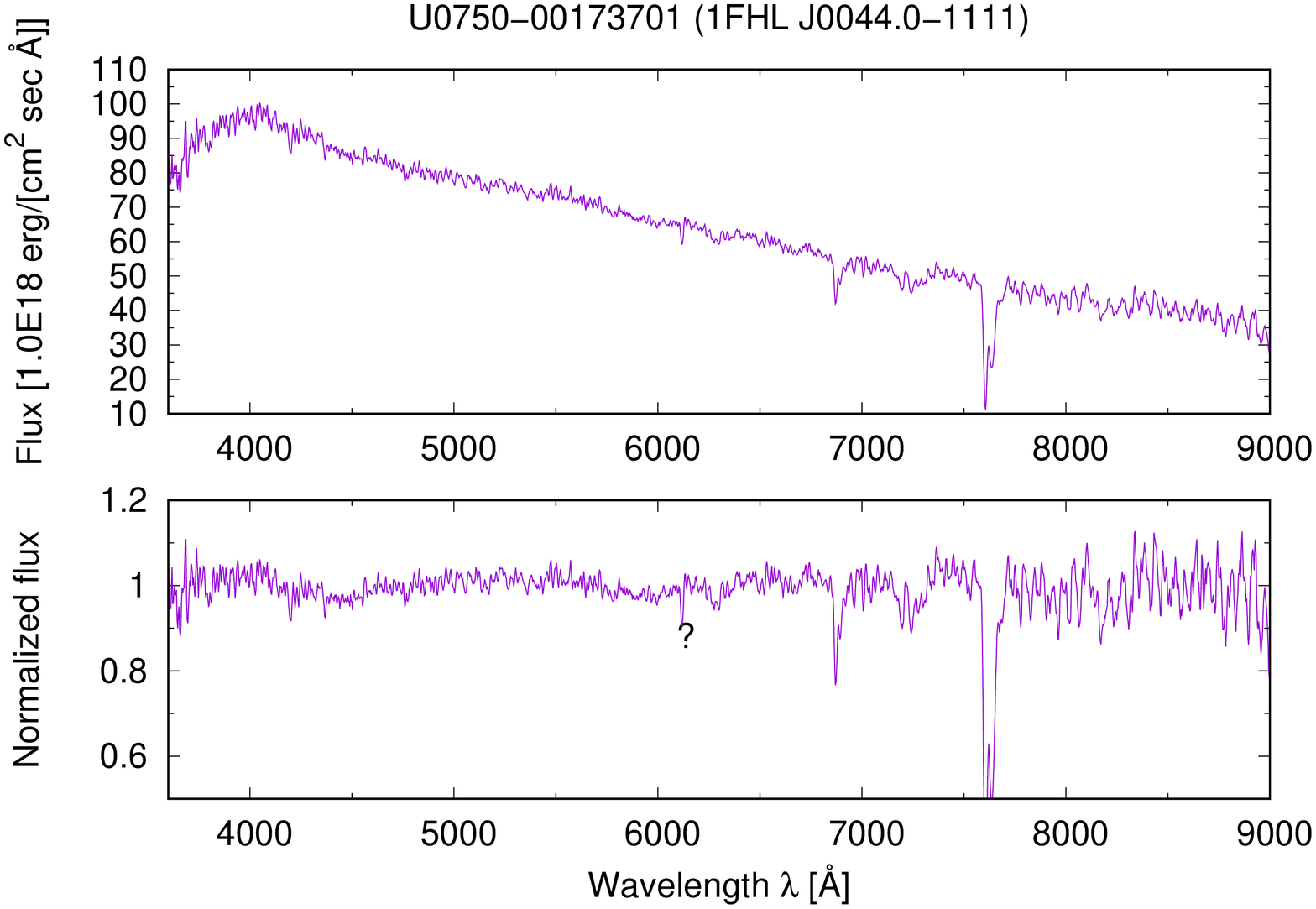}
     \includegraphics[angle=270,width=0.463\hsize]{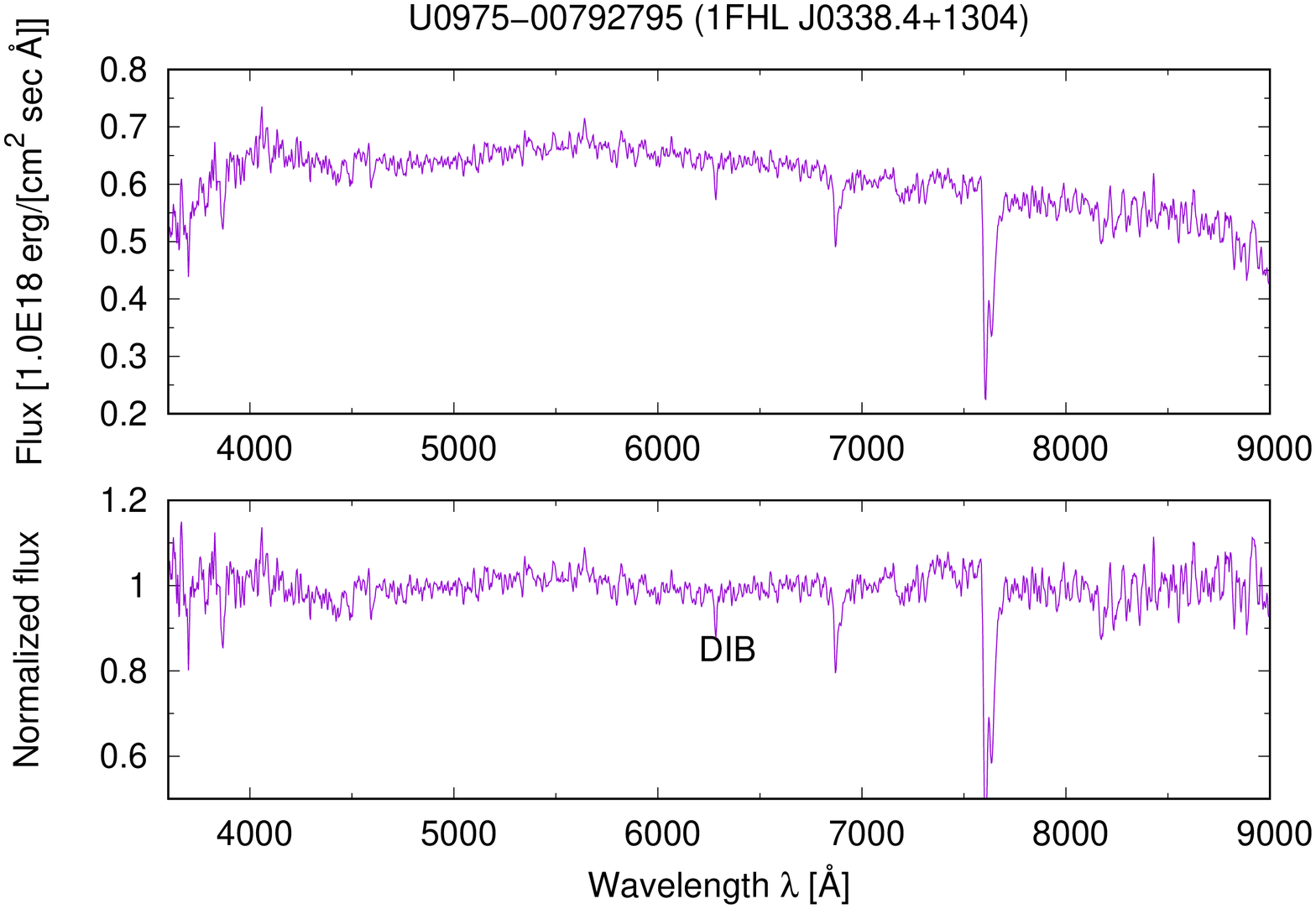}
     \includegraphics[angle=270,width=0.463\hsize]{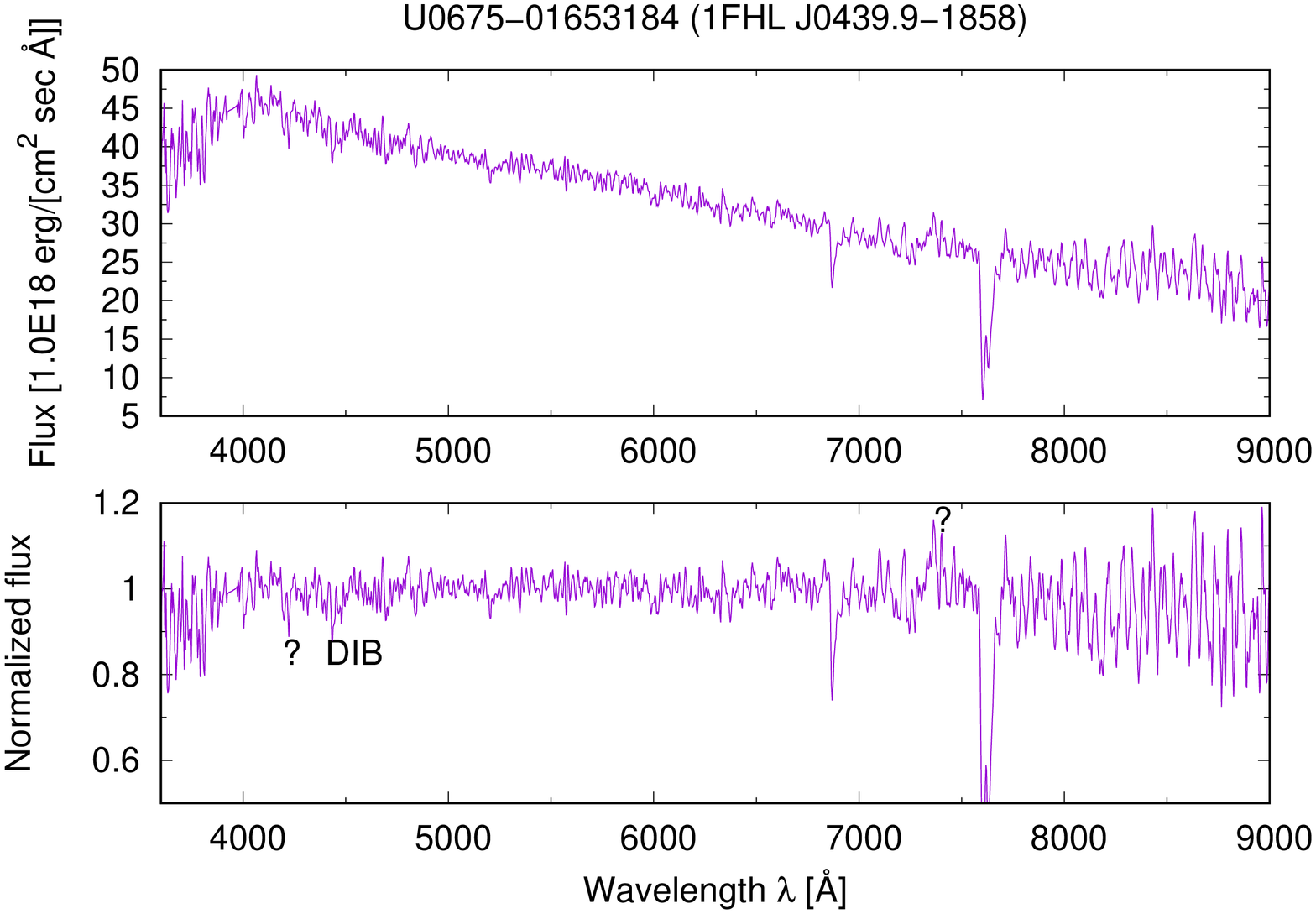}
     \includegraphics[angle=270,width=0.463\hsize]{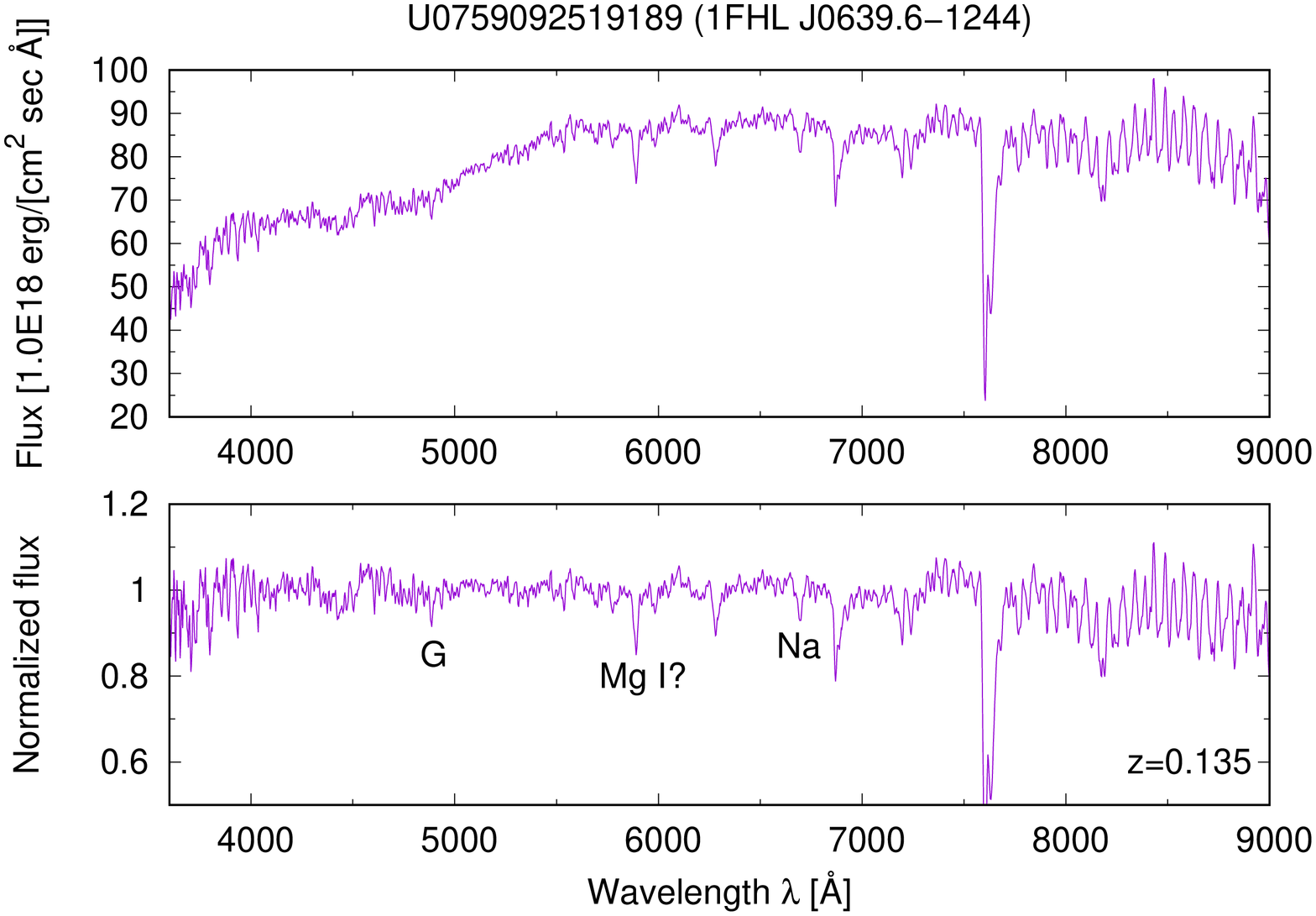}
     \includegraphics[angle=270,width=0.463\hsize]{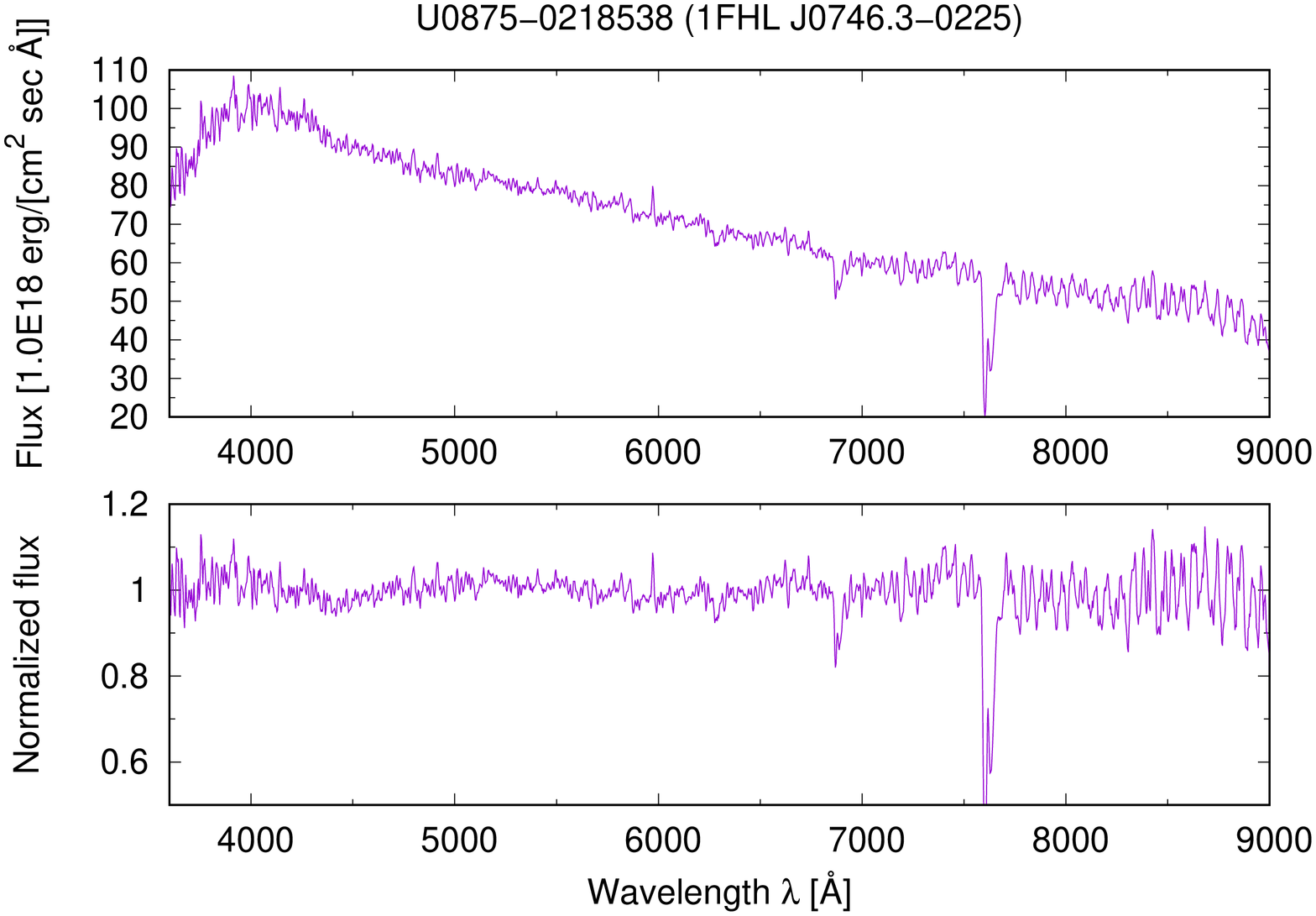}
     \includegraphics[angle=270,width=0.463\hsize]{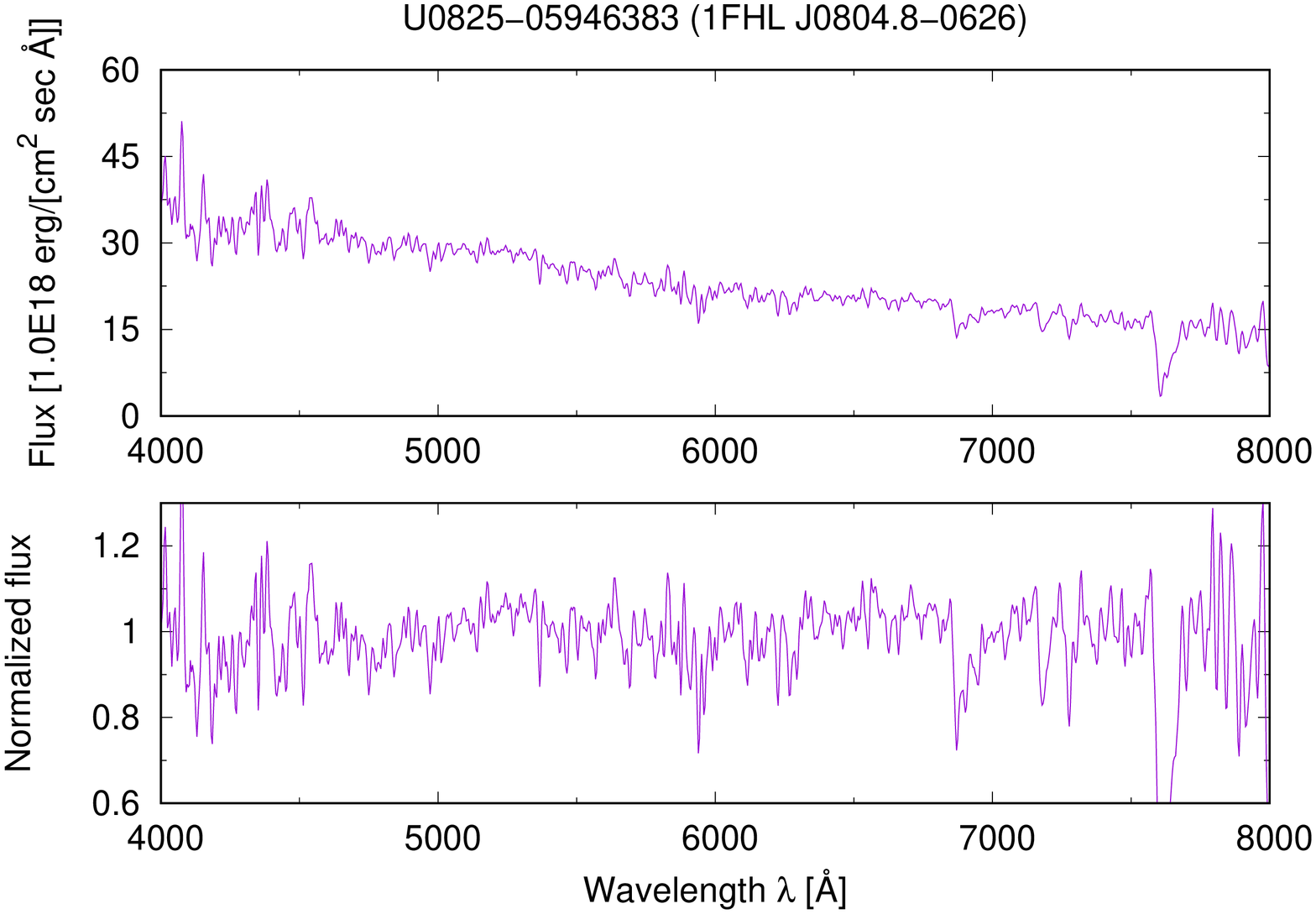}
     \includegraphics[angle=270,width=0.463\hsize]{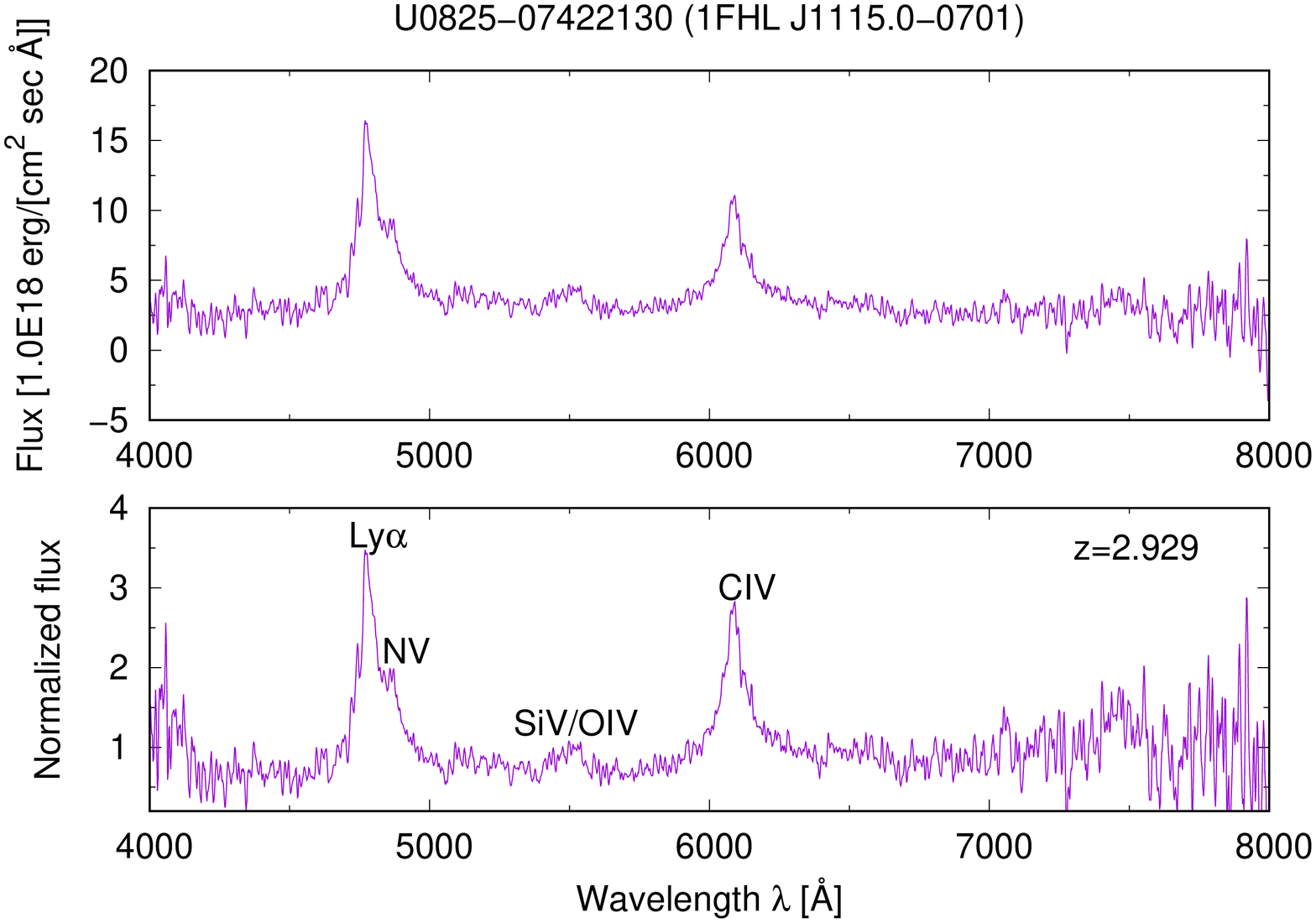}
     \includegraphics[angle=270,width=0.463\hsize]{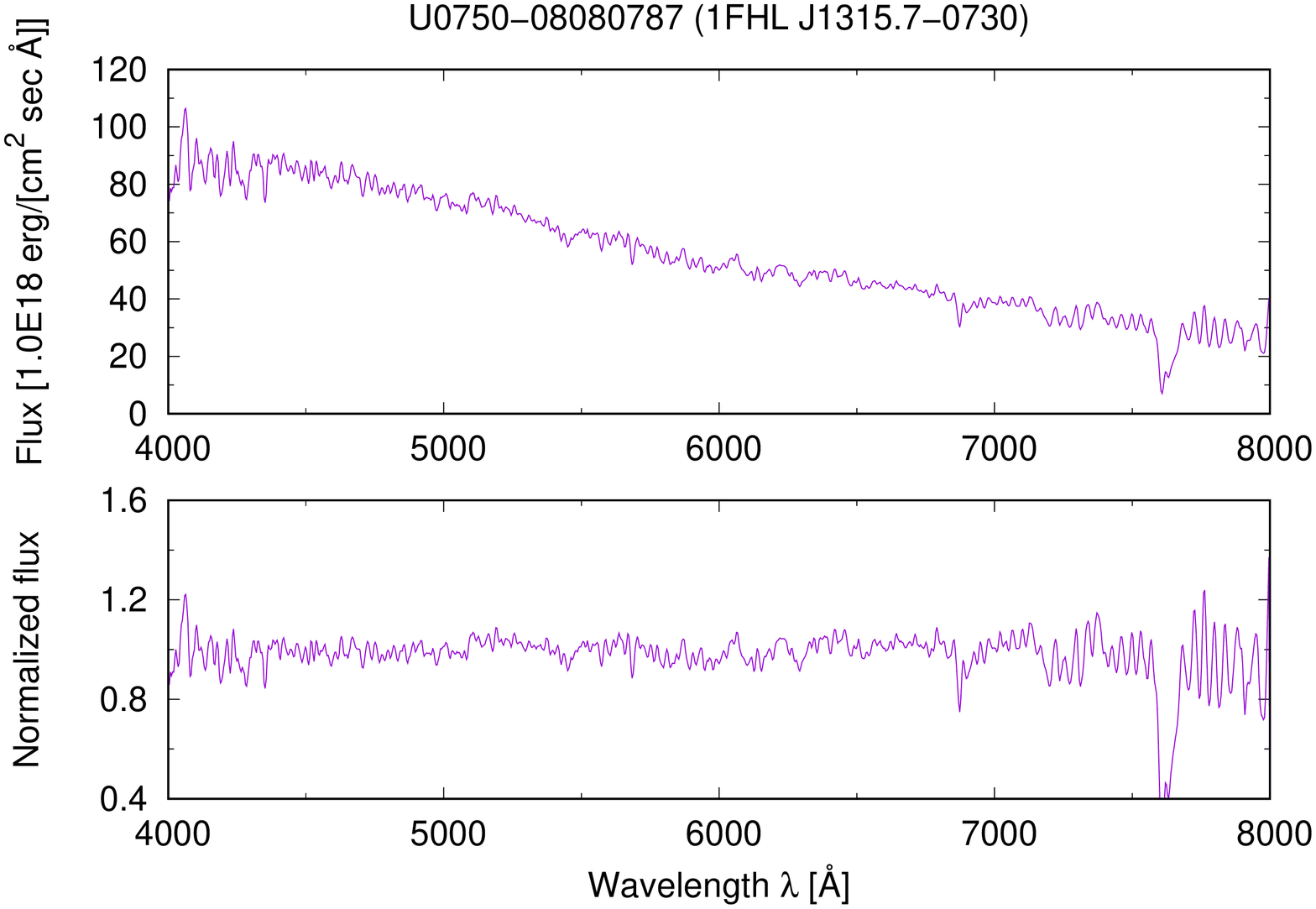}
   
    \caption{Optical spectra obtained for the whole sample presented in this paper. Upper panels show the observed spectra, while lower panels show the spectra with normalised flux. Absorption lines or bands present at 686.9\,nm, $\sim$718.6\,nm, and 760.5\,nm are telluric. Absorption lines present at 589.0\,nm and 589.6\,nm correspond to the NaI doublet from the interstellar medium, although in the case of 1FHL\,J0639.6-1244 it could possibly be superimposed on the MgI line at $z=0.135$. Lines marked 'DIB' correspond to diffuse interstellar bands, while those marked with a question mark are hard to identify because they are on the edge of detection and because of the lack of other lines to obtain a redshift value. Sources are given with their USNO designator, while the proposed 1FHL counterpart is given in parenthesis.}
    \label{fig1}
    
   \end{figure*}

\addtocounter{figure}{-1}
   \begin{figure*}[!htp] \centering
   
     \includegraphics[angle=270,width=0.47\hsize]{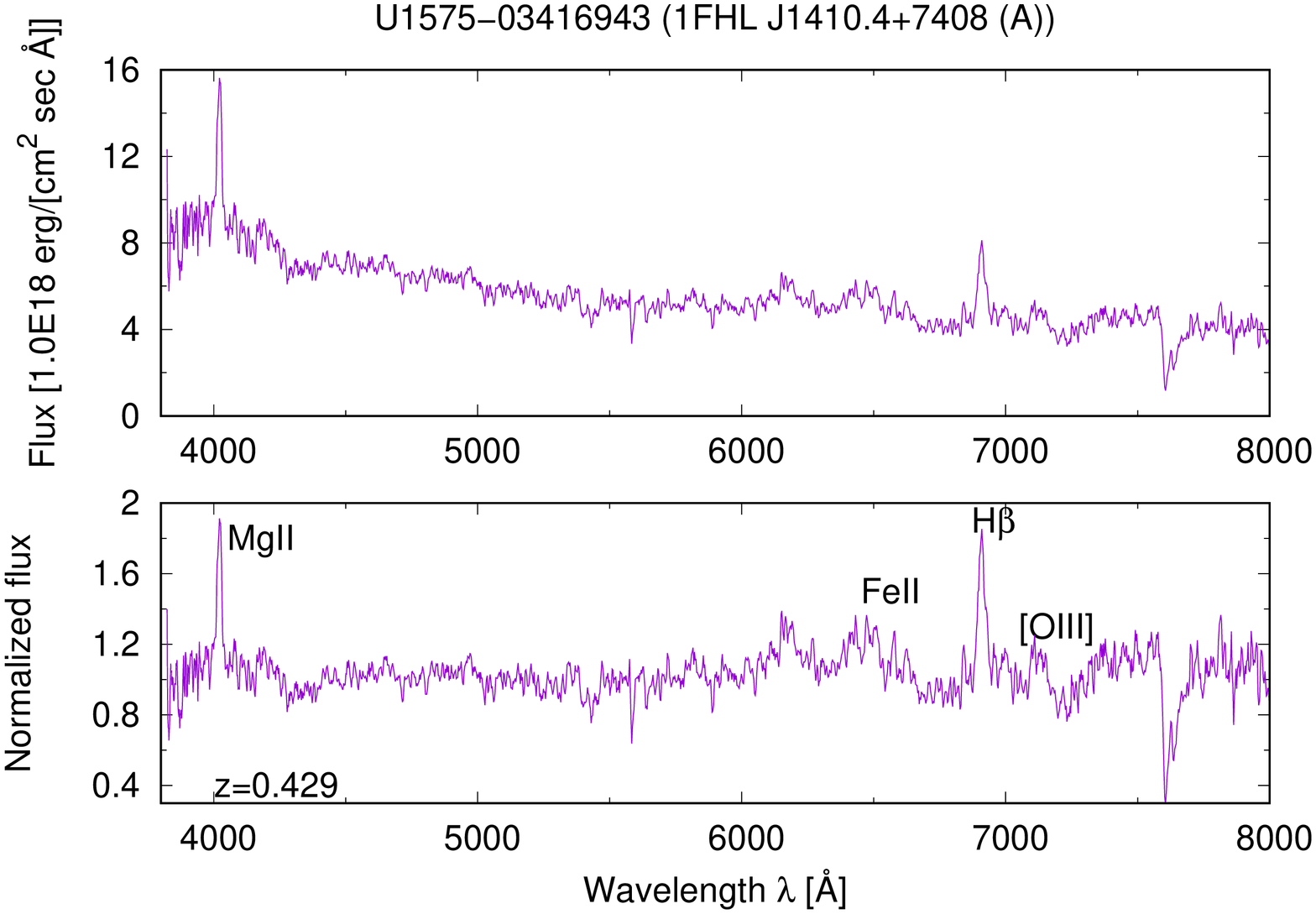}
     \includegraphics[angle=270,width=0.47\hsize]{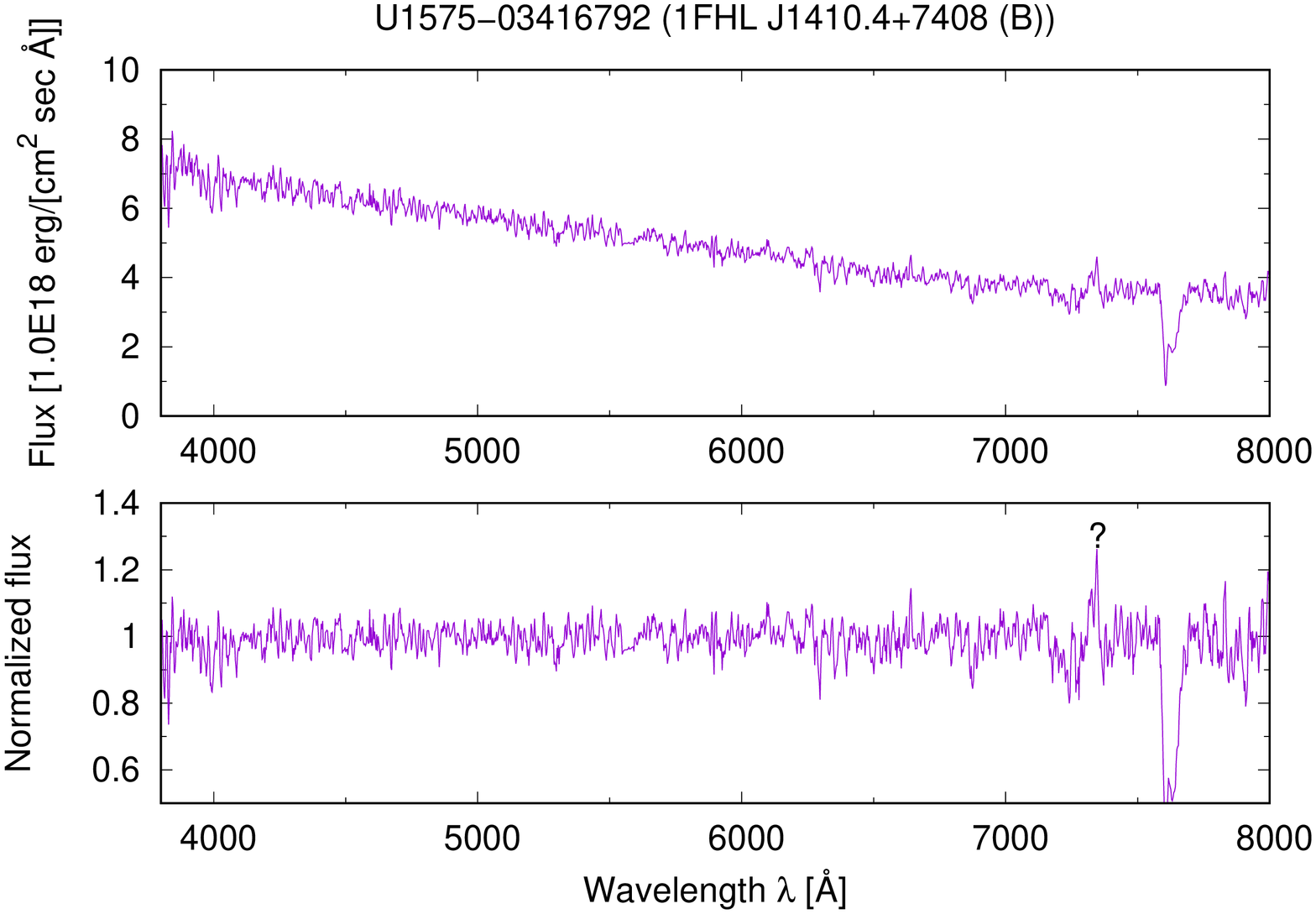}
     \includegraphics[angle=270,width=0.47\hsize]{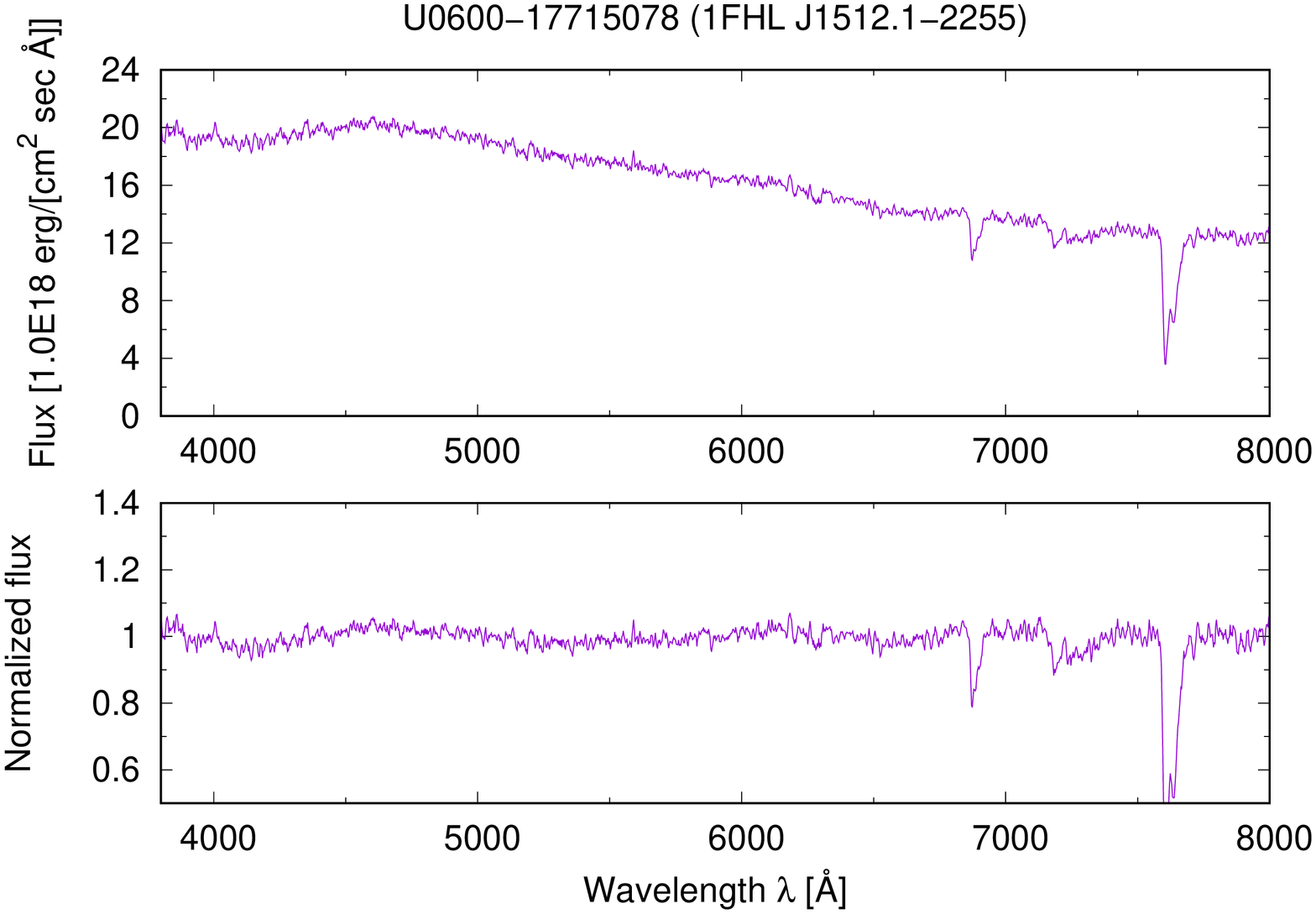}
     \includegraphics[angle=270,width=0.47\hsize]{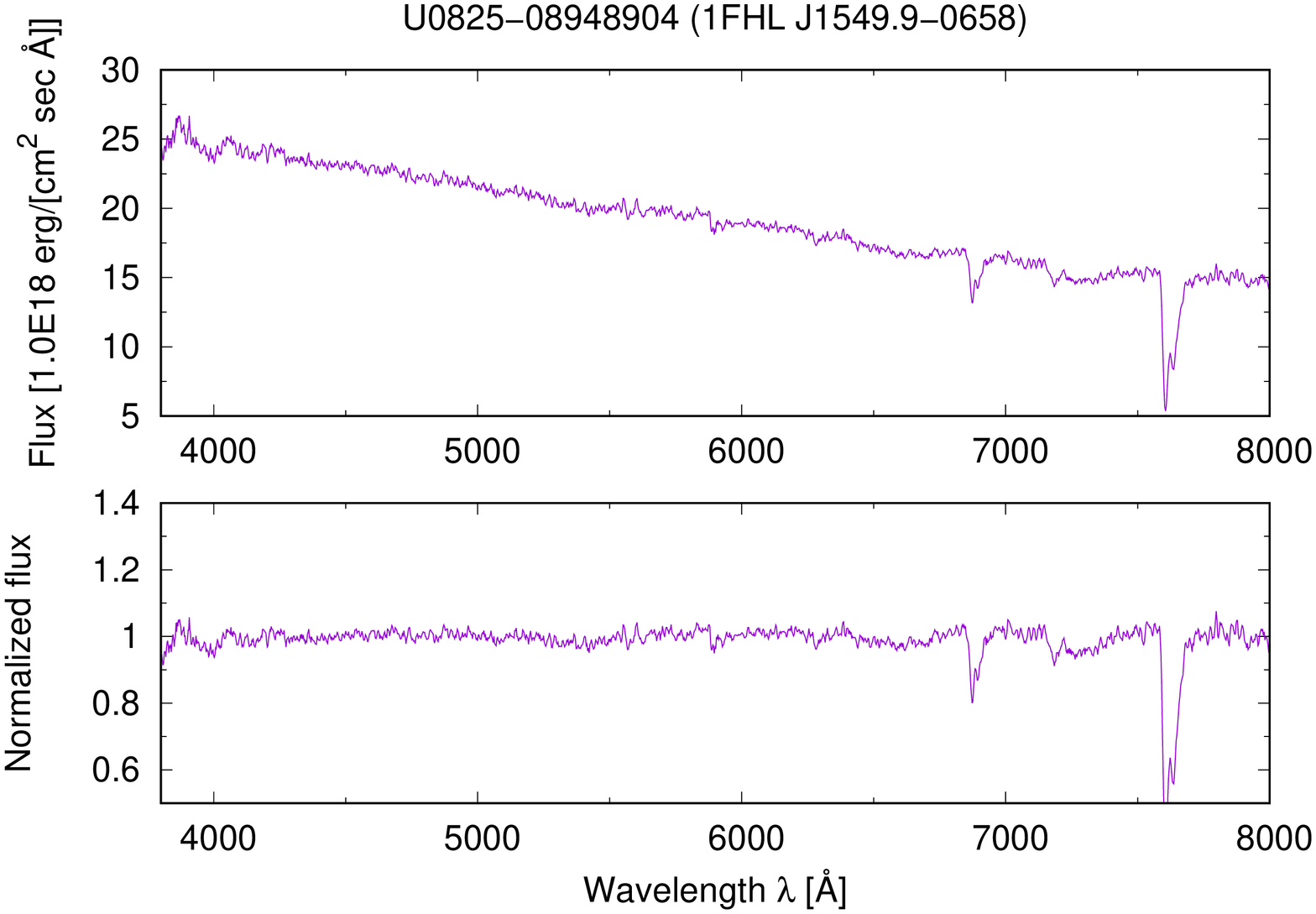}     
     \includegraphics[angle=270,width=0.47\hsize]{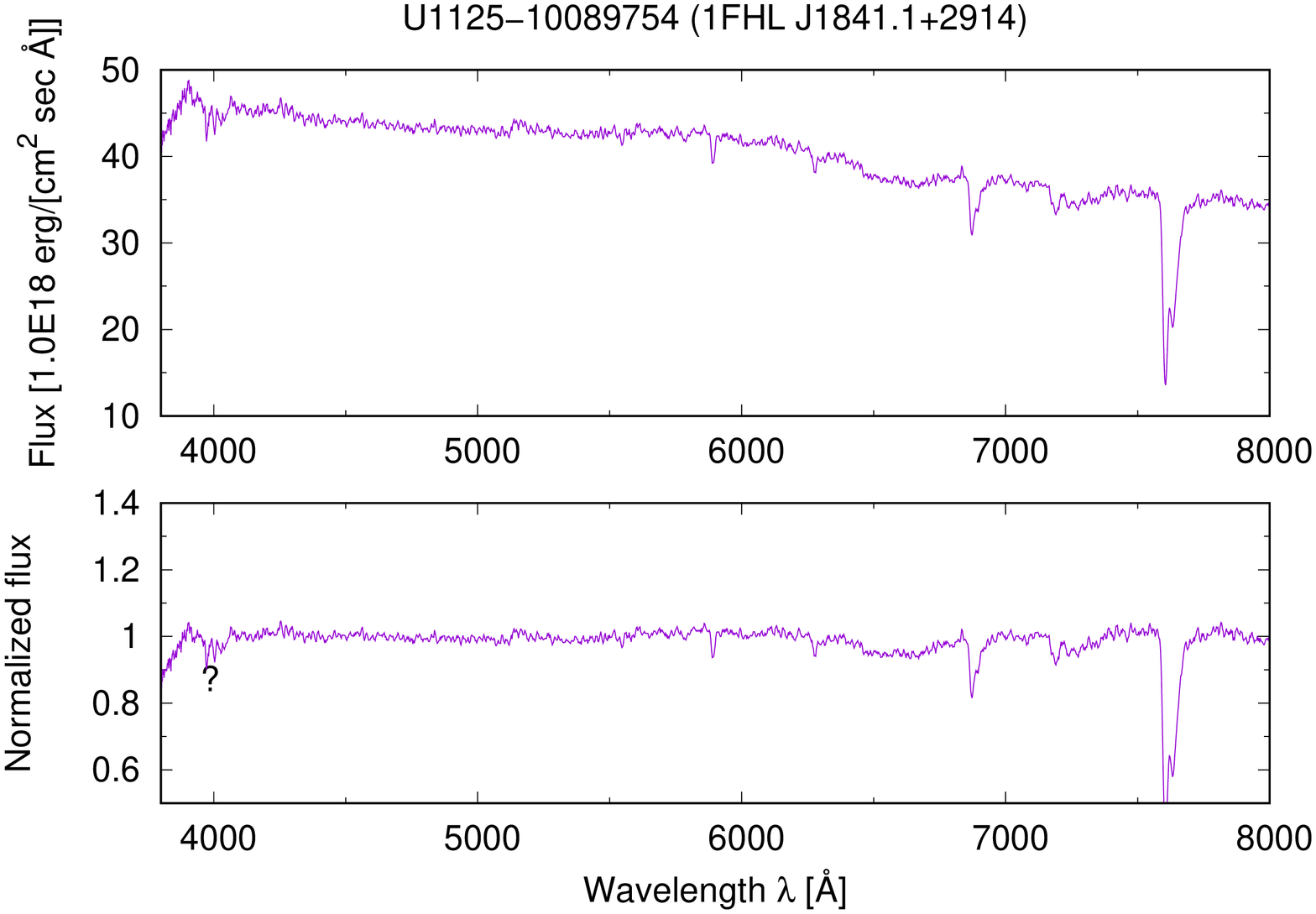}
     \includegraphics[angle=270,width=0.47\hsize]{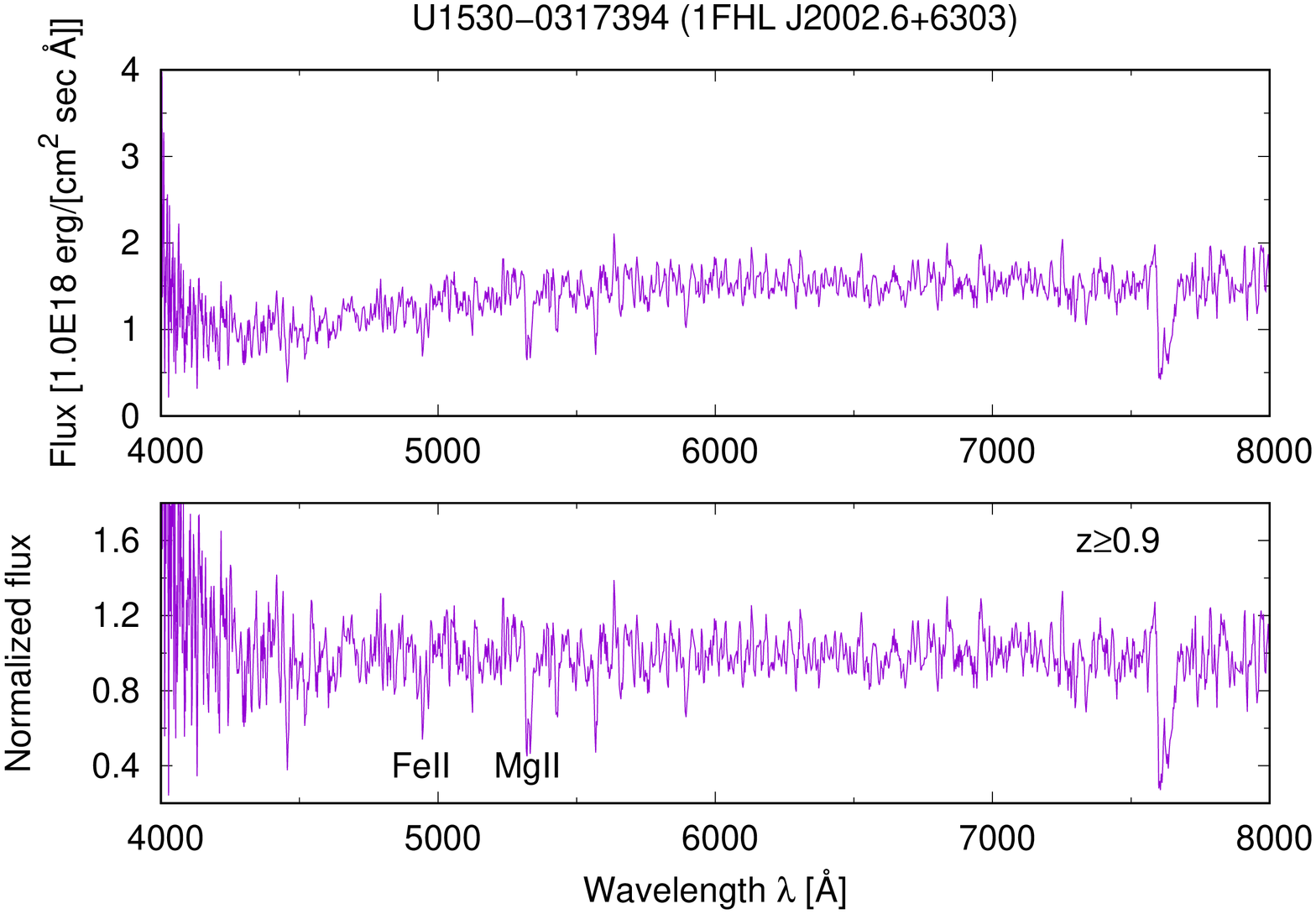}
     
    \caption{(Continued).}
    \label{fig2}
    
   \end{figure*}


   \begin{table*}[!htp]
   
   \caption{Nature of each of the observed optical counterpart candidates for 1FHL sources.} 
   \label{tab2} 
   \centering
   
   \setlength{\tabcolsep}{6pt}
   \begin{tabular}{cccccc}
   \hline\hline
   \\
    USNO designator			& Features & EW [\AA]      & Flux   & Redshift & Class \\ 
1FHL association (Distance) &&&&& \\
   (1) & (2) & (3) & (4) & (5) & (6) \\
   
   \hline\hline

U0750-00173701							&   --        & --          &   --         &  --      & BL Lac \\
\rm{1FHL\,J}0044.0-1111 (5.7)			&   --        & --          &   --         &  --      &  \\
\hline
U0975-00792795							&   --        & --   	     &   --         &  --      & BL Lac \\
\rm{1FHL\,J}0338.4+1304	(2.5)			&   --        & --          &   --         &  --      &  \\
\hline
U0675-01653184							&   --        & --          &   --         &  --      & BL Lac \\
\rm{1FHL\,J}0439.9-1858	(2.9)			&   --        & --          &   --         &  --      &  \\
\hline							
U0750-02519189  						& G           &1.0$\pm$0.6  & -8$\pm$-4      & 0.135$\pm$0.001  & BL Lac \\
\rm{1FHL\,J}0639.6-1244 (10.7)     		& Na          &2.3$\pm$1    & -21$\pm$-10    &         		 &        \\
\hline			         		
U0875-0218538							&   --        & --          &   --         &  --      & BL Lac \\
\rm{1FHL\,J}0746.3-0225	 (1.7)			&   --        & --          &   --         &  --      &  \\
\hline
U0825-05946383							&   --        & --          &   --         &  --      & BL Lac \\
\rm{1FHL\,J}0804.8-0626	 (2.1)			&   --        & --          &   --         &  --      &  \\
\hline
U0825-05946383							& Ly$\alpha$& 312$\pm$28     & 95$\pm$10   & 2.929$\pm$0.003     &  QSO   \\
\rm{1FHL\,J}1115.0-0701	 (3.2)			& NV 		& 262$\pm$49     & 76$\pm$12   &          &        \\
										& SiV/OIV   & 79$\pm$14      & 21$\pm$3    &          &        \\
										& CIV   	& 264$\pm$31     & 82$\pm$6    &          &        \\
\hline							
U0750-08080787							&  --       & --             &     --      &  --      & BL Lac \\
\rm{1FHL\,J}1315.7-0730	 (3.3)			&  --  		& --             &     --      &  --      &  \\
\hline
					U1575-03416792	   	& MgII      & 17$\pm$6       & 15$\pm$4    & 0.429$\pm$0.001     & NLS1  \\
\rm{1FHL\,J}1410.4+7408 A (4.4)			& H$\gamma$ & 18$\pm$9       & 9$\pm$4     &          &        \\
										& H$\beta$  & 35$\pm$14      & 14$\pm$5    &          &        \\
										& [OIII]    & --             &      -- 	   &          &        \\
\hline							 
					U1575-03416943	  	&   --        & --      	&   --         &  --      & BL Lac \\
\rm{1FHL\,J}1410.4+7408 B (3.3)			&   --        & --          &   --         &  --      &  \\
\hline
				U0600-17715078			&   --        & --          &   --         &  --      & BL Lac \\
\rm{1FHL\,J}1512.1-2255 (1.0)			&   --        & --          &   --         &  --      &  \\
\hline	
						U0825-08948904 	&   --        & --    	 	&   --         &  --      & BL Lac \\ 
\rm{1FHL\,J}1549.9-0658  (1.4)			&   --        & --          &   --         &  --      &  \\
\hline 
					U1125-10089754		&   --        & --          &   --         &  --      & BL Lac \\
\rm{1FHL\,J}1841.1+2914 (5.6)			&   --        & --          &   --         &  --      &  \\
\hline
						U1530-0317394	& FeII        & 11$\pm$7    & -1.5$\pm$-0.9  & $\ge$0.9 & BL Lac \\
\rm{1FHL\,J}2002.6+6303	  (1.1) 		& MgIIa       & 10$\pm$3    & -1.6$\pm$-0.6  &          &        \\
										& MgIIb       & 7$\pm$2     & -1.1$\pm$-0.4  &          &        \\
\hline   
\hline

\multicolumn{6}{c}{Notes: The units for all the reported flux densities are $\rm{1\times 10^{-16}\cdot erg \cdot cm^{-2} \cdot sec^{-1} \cdot \AA^{-1}}$.}\\
\multicolumn{6}{c}{The equivalent width (EW) is given in the observer's frame. Emission lines are given}\\
\multicolumn{6}{c}{as positive flux values, and absorption lines as negative flux values. The distance}\\
\multicolumn{6}{c}{between the USNO source and the 1FHL centroid is given in arcseconds.}\\
   \end{tabular}
   \end{table*}
   \normalsize

In 12 out of 14 cases, the spectra resulted in non-thermal continua. Moreover, no intrinsic features were present in 10 out of 14 objects. Both are typical characteristics of blazar spectra. In all cases in which some features are found, a redshift (or at least a lower limit to it) was derived, in addition to obtaining equivalent widths and fluxes for all lines, in order to determine the nature of each source. Results from our analysis can be found in Table \ref{tab2}, where we report in column 1 the USNO source name along with the name of the proposed 1FHL counterpart and the distance between them, in column 2 the emission and/or absorption lines found (if any), in columns 3 and 4 their measured equivalent widths and fluxes, in column 5 the derived redshift (if any), and in column 6 the classification of the source. Further details are shown in the next sections.

It is worth mentioning that, in the cases of 1FHL\,J1115.0-0701 and 1FHL\,J0804.8-0626, the correlation with X--ray data showed only one source inside the $\gamma$--ray positional error area, while in optical wavelengths (as seen in the USNO plates, with a limiting magnitude of $V\approx21$ mag) two objects could be found within the X--ray error circle. In both cases, the other object was also analysed and ruled out because of its star--like spectrum, i.e. showing a thermal continuum, no emission lines, and a variety of absorption lines potentially associated with stellar processes (for instance, the Balmer series) at redshift zero. A different case is that of fields 1FHL\,J1410.4+7408 A and B, which are potentially associated with the same source in the 1FHL catalogue but for which two X--ray objects were found within the $\gamma$--ray error ellipse \citep{Landi15a} and, consequently, two putative optical counterparts could be potentially associated with this $\gamma$--ray source. This case will be discussed in Section 5.2.

Once confirmed as potential counterparts (i.e. after discarding all the sources from which no high-energy emission is expected, as for example stars), we improved their equatorial coordinates by searching for detected objects in the 2MASS \citep{Skrutskie06} catalogue, which provides positions with uncertainties of less than 0.1 arcsec. Only four of them were not found in this catalogue: the optical sources potentially associated with 1FHL\,J1410.4+7408A, 1FHL\,J1410.4+7408B, 1FHL\,J1549.9-0658, and 1FHL\,J1841.1+2914. Nevertheless, the first three were found in the USNO-A2.0 catalogue \citep{Monet98}, and the last one in the USNO-B1.0 catalogue \citep{Monet03}, which provide an accuracy of 0.2 arcsec.

Source details are given in the following subsections.

\subsection{BL Lacs}

Out of the 14 optical sources observed, we were able to classify 12 as blazars of BL Lac class: These are our associations with Fermi sources 1FHL\,J0044.0-1111, 1FHL\,J0338.4+1304, 1FHL\,J0439.9-1858, 1FHL\,J0639.6-1244, 1FHL\,J0746.3-0225, 1FHL\,J0804.8-0626, 1FHL\,J1315.7-0730, 1FHL\,J1410.4+7408 B, 1FHL\,J1512.1-2255, 1FHL\,J1549.9-0658, 1FHL\,J1841.1+2914, and 1FHL\,J2002.6+6303. Indeed, all the sources show a non--thermal, power--law, intrinsically blue continuum, and no apparent intrinsic emissions or absorptions, with the exception of U0750-02519189 (associated with 1FHL\,J0639.6-1244), in which its host galaxy contribution is visible (meaning it is a blazar of the BZG type, as described by the Roma-BZCAT catalogue in \citet{Massaro15z}), showing Na and G-band absorptions at a redshift $z=0.135\pm 0.001$. This, alongside a lower limit for the redshift of our association for 1FHL\,J2002.6+6303, U1530-0317394 ($z\ge 0.9$), obtained from the detection of intervening FeII and MgII absorptions, is the only value for $z$ we were able to derive from the spectra of this BL Lac subsample. 

In the case of the optical association of 1FHL\,J0804.8-0626, there are two optical sources inside the X--ray error box, according to the USNO plates. Both of them were observed and analysed. The faintest one (at optical position $8^h04^m58\fs48^s, -6^{\circ}24'21\farcs1$) showed a normal G-type star spectrum, thus discarding any possibility of potential association with the high--energy emitting source. The coordinates published in Table \ref{tab1} are thus those of the BL Lac conclusively associated through optical spectroscopy, which is the WISE source suggested by \citet{Landi15c} and which we associate with the $\gamma$--ray source.

\subsection{U1575-03416943}

This source potentially associated with 1FHL\,J1410.4+7408 A shows clear emission lines (MgII, H$\delta$, H$\gamma$, H$\beta$, and [OIII]) at a common redshift $z=0.429\pm 0.001$. Given that the velocities associated with the emission of the H$\beta$ line are around $1450\, \rm{km/s}$, and that the ratio between the fluxes of emission lines $[\rm{OIII}]$ and $\rm{H}\beta$ is $\le 0.5$, we conclude that this object is a narrow line Seyfert 1 galaxy \citep[NLS1,][]{Osterbrock85,Goodrich89}.

\subsection{U0825-05946383}

The field associated with 1FHL J1115.0-0701 presented two optical sources within the X--ray positional uncertainty box, according to the USNO plates. In this case, again, both spectra were analysed, and we could discard one of them on the basis of typical stellar features (in particular, we classified it as a K-type star, at position $11^h 15^m 15\fs3^s,-07^{\circ}01'26\farcs9$).

The spectrum of the other optical source shows strong, luminous emission lines for Ly$\alpha$, NV, SIV, and CIV, at the high redshift value of $z=2.929\pm 0.003$; these characteristics are typical of a high--redshift quasar. However, its potential association with the 1FHL source is not ironclad (see Section 5.3).

\section{Discussion}

In this section we analyse in detail the spectral characteristics of the results obtained for the 14 objects we spectroscopically associated in this work. In particular, we discuss general properties in subsets divided by class of object: BL Lacs (12 objects), NLS1 (1 object), and quasars (1 object).

\subsection{BL Lacs}

In order to analyse the emission processes involved, we built a plot of spectral indices as shown in \citet{Abdo10c}, which is useful to easily spot the synchrotron peak for each object. To this end, and following \citet{Masetti13}, we searched for the X--ray fluxes of the sources in our sample as measured with XRT or ROSAT, corrected from Galactic absorption with PIMMS \citep{Mukai93} using the Galactic $N_H$ values given by \citet{Landi15a}, when available, or those given by \citet{Kalberla05}. We also retrieved their $R$ magnitudes from the USNO catalogues, from which we derived absorption-corrected fluxes using the absorption maps from \citet{Schlegel98}, the reddening law of \citet{Cardelli89}, and the total-to-selective extinction ratio of \citet{Rieke85}; with the conversion factor of \citet{Fukugita95} we then rescaled the flux values to 500\,nm using the same procedure given by \citet{Masetti13}. Furthermore, we obtained their radio flux densities at 1.4GHz, when available, from the NVSS catalogue \citep{Condon98} and rescaled them to 5\,GHz assuming a radio flat spectral shape \citep{Begelman84} in order to use the same relationship given in \citet{Abdo10c}.

\begin{figure}[!htp]
    \centering
   
     \includegraphics[angle=270,scale=0.45]{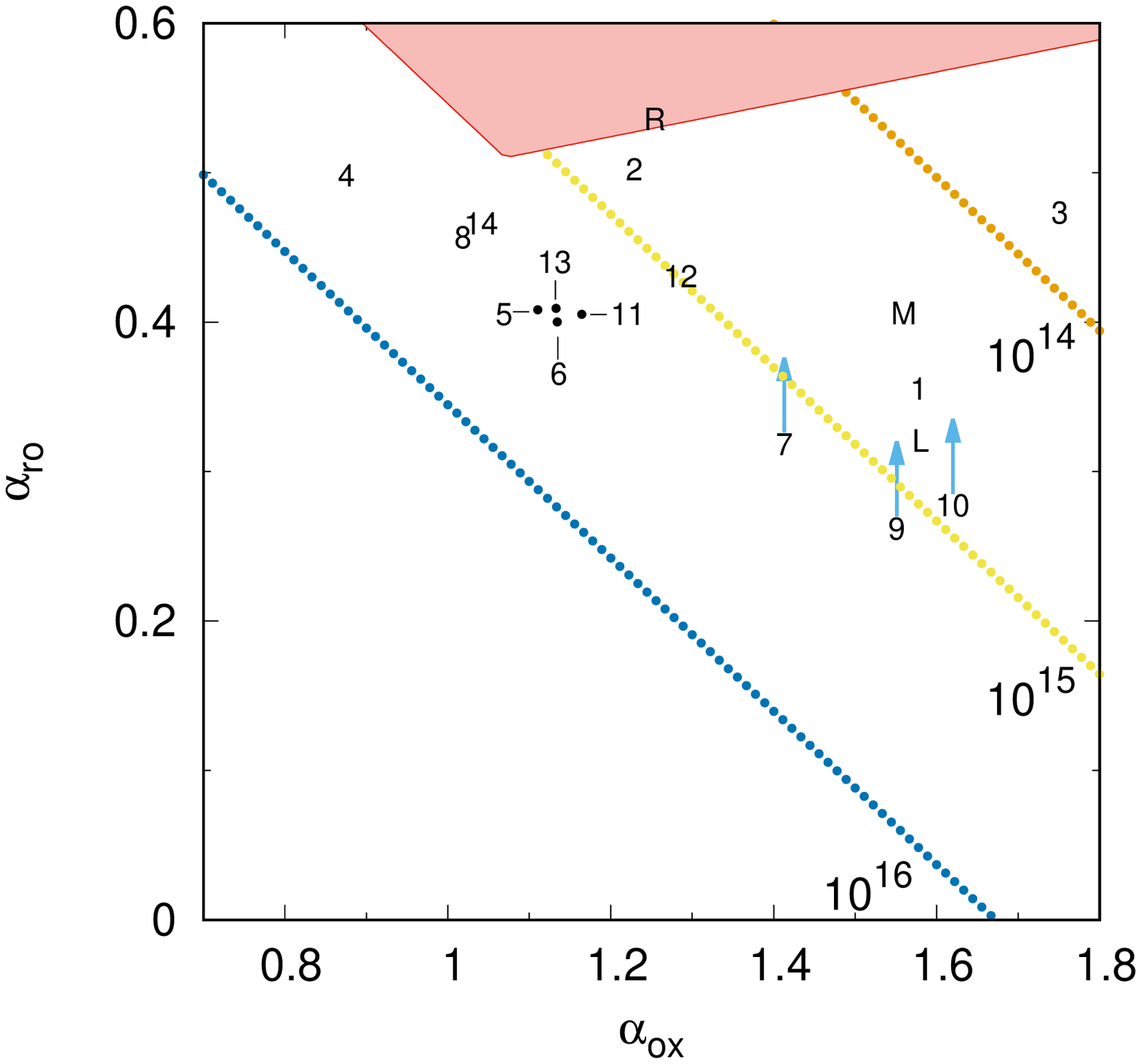}
     
    \caption{Analysed sources in the spectral indices $\alpha_{ox}-\alpha_{ro}$ plane. Lower limits are indicated with an arrow. Numbers refer to the values presented in Table \ref{tab1}, while letters L, M, and R refer to the objects in \citet{Landi15a,Landi15c} associated by \citet{Landoni15}, \citet{Massaro15}, and \citet{Ricci15}, respectively. Sources 5, 6, 11 and 13 are shown as points for the sake of clarity. The red shaded area indicates the region where the relationship describing synchrotron peak values changes its functional form, as explained in \citet{Abdo10c}.}
    \label{fig3}
   \end{figure}

With the radio, optical, and X-ray absorption-corrected fluxes we were able to obtain spectral indices $\alpha_{ox}$ from X--ray to optical and $\alpha_{ro}$ from optical to radio frequencies. In Figure \ref{fig3}, we included all the sources from this sample, numbered in order of right ascension (see Table \ref{tab1}), in a $\alpha_{ox}-\alpha_{ro}$ plot \citep{PadovaniGiommi95,Abdo10c}. In dashed lines, we indicate the location of synchrotron peaks at low ($10^{14}\rm{Hz}$), intermediate ($10^{15}\rm{Hz}$), and high ($10^{16}\rm{Hz}$) frequencies. Eight BL Lacs in our sample have their synchrotron peaks at a frequency higher than $10^{15}\rm{Hz}$, which are likely candidates to be detected at TeV energies \citep{Massaro08}. It is important to highlight that the BL Lac associated with 1FHL\,1410.4+7408 B did not show any radio emission, which is why we used the detection threshold from the NVSS survey ($2.5\rm{mJy}$) as upper limit to its radio flux density. The resulting lower limit to $\alpha_{ro}$ is indicated in the plot with an arrow.

For completeness, we also included the recently studied optical objects associated with 1FHL\,J1129.2-7759, 1FHL\,J1328.5-4728, and 1FHL\,2257.9-3644 which were confirmed as BL Lacs by \citet{Massaro15} (for which we found an intermediate synchrotron peak, marked with an M in Figure \ref{fig3}), \citet{Ricci15} (which shows a low synchrotron peak, marked with an R) and \citet{Landoni15} (intermediate synchrotron peak, marked with an L), respectively. These objects are also part of the sample selected by \citet{Landi15c} and \citet{Landi15a}. In addition, we added the two non-BL Lac objects from our sample, the potential associations with 1FHL\,J1410.4+7408 A (the NLS1 presented in Section 4.2) and 1FHL\,J1115.0-0701 (the quasar) just to present the whole sample in one plot, although it is not possible to compare these sources with BL Lacs given that this kind of objects generally present a thermal emission component which cannot be easily separated from the non-thermal one. Neither one presents radio emission, as seen in Figure \ref{fig3}, so also in this case we used the NVSS threshold value to determine a lower limit for $\alpha_{ro}$.

\subsection{The case of 1FHL\,J1410.4+7408}

As 1FHL\,J1410.4+7408 is potentially associated with two different X--ray emitting and optically peculiar objects according to \citet{Landi15a}, it is not clear which of them is responsible for the detected $\gamma$--ray emission. 


Given that the spectral index of the $\gamma$--ray source, according to the 1FHL catalogue, is $\alpha_{\gamma}=2.65$, its counterpart is more likely a flat spectrum radio quasar (FSRQ) than a BL Lac \citep{Ackermann15}. We were not able to find any radio counterpart association in public surveys for the NLS1 potentially associated with 1FHL\,J1410.4+7408\,A. Although NLS1 have been indicated as responsible for $\gamma$--ray as well as X--ray emission \citep[e.g.][and references therein]{Abdo09,Foschini15}, the fact that it is not detected at radio bands brings up the possibility that this association is the product of a contamination of the sample due to the relative width of the Fermi positional error boxes. Only radio loud NLS1 have been detected in high energies.

Likewise, the BL Lac object probably associated with 1FHL\,J1410.4+7408\,B also does not show radio emission. It is important to note that, if confirmed, the BL Lac object 1FHL\,J1410.4+7408\,B would be one of the very few radio quiet $\gamma$--ray emitting BL Lac objects identified to date. Similar recent cases can be found in \citet{Paggi14} and in \citet{Ricci15}.

To be conservative, it is thus safe to say that it is still not clear which of the two sources is the actual $\gamma$--ray emitter, or that the two objects are possibly contributing to the total $\gamma$--ray flux detected with Fermi. However, given the above considerations, it is more likely that the counterpart to this 1FHL $\gamma$--ray source is the BL Lac object associated with 1FHL\,J1410.4+7408\,B.

Regarding the NLS1 object associated with 1FHL\,J1410.4+7408\,A, a central black hole mass value can be estimated through measuring the FWHM and flux of the $H_{\beta}$ line \citep{Kaspi00,Wu04} corrected for foreground galactic absorption. This allows us to infer a mass of $\sim5\times10^{6}\rm{M}_{\sun}$ for the black hole.

\subsection{The case of 1FHL\,J1115.0-0701}

We classified the optical counterpart of the X--ray source found within the 1FHL\,J1115.0-0701 positional uncertainty ellipse as a high--redshift quasar, with $z=2.929\pm 0.003$. This value, in turn, allows us to estimate a luminosity distance of $\sim24.7$ Gpc, assuming $\rm{H_0}=70.0$, $\Omega_{\rm{m}}=0.3$ and $\Omega_{\rm{\Lambda}}=0.7$. Following \citet{Park13}, we measured the flux and width of the CIV emission line together with the flux level of the continuum at 135\,nm (rest--frame), both corrected for foreground galactic absorption, to obtain an estimate for the mass of its central black hole. This resulted in $4.6\times10^{9}\rm{M}_{\sun}$, which is within the range of expected black hole masses for this kind of AGN \citep{Vestergaard06}. Moreover, the above distance estimate implies an X--ray luminosity of $L_X=7.9\times10^{45}\rm{erg/s}$ in the 2--10 keV band, whereas the black hole mass corresponds to an Eddington luminosity value  $L_{\rm{Edd}}=5.5\times10^{47}\rm{erg/s}$. Adopting a correction factor $C_X=15.8$ to obtain the X--ray bolometric luminosity \citep{Ho09}, we find an Eddington ratio of $L_{\rm{X}}/L_{\rm{Edd}} = 0.2$. Assuming this quasar is the real counterpart for the 1FHL source, its $\gamma$--ray luminosity results in $L_{\gamma}=3.2\times10^{47}\rm{erg/s}$. This value, although rarely reached, is within the range expected for $\gamma$--ray emitting FSRQs \citep{Cavaliere02}.

However, \citet{Petrov13}, \citet{Massaro13}, and \citet{Schinzel15} proposed a potential association of this Fermi source with a radio object (NVSS\,J111511--070238) located at a distance $\sim90\,\rm{arcsec}$ from the X--ray source found by \citet{Landi15a}. The radio source NVSS\,J111511--070238 is located at a distance of $\sim$2.5\,arcmin from the 1FHL source, while the X--ray source lies at a distance of $\sim$3.2\,arcmin from the latter. These two objects are not positionally consistent with each other. Therefore, this suggests that there may be two AGN within the Fermi error ellipse, a radio emitting one and an X--ray emitting one, which is the one we classify as a quasar.

In an attempt to discard one of the two proposed counterparts, we searched for archival multiwavelength data for both sources. We found no radio emission at the position of the X--ray quasar, suggesting that the object is possibly radio quiet and/or too cosmologically distant to be detected in the NVSS. However, this non--detection does not completely rule out the quasar as the real counterpart. Figure 14 of \citet{Abdo09b} suggests a connection between radio luminosities and the $\gamma$--ray spectral indices obtained with the whole energy band at which Fermi/LAT works (i.e. 20 MeV to 300 GeV). If this object falls on the faint side of the connection ($\sim 1\times 10^{42}\,\rm{erg/s}$), shallow radio surveys are not able to detect any emission: indeed, at a redshift $z=2.929$ that luminosity would correspond to a flux density of $\sim 1\,\rm{mJy}$, which is well below the detection threshold of the NVSS ($2.5\,\rm{mJy}$).

Moreover, given that the spectral index across the whole Fermi/LAT energy range \citep[$\alpha_{\gamma}=2.11$,][]{Acero15} for this $\gamma$--ray source is an intermediate value between those of each kind, we cannot determine whether it is a BL Lac or a FSRQ.

On the other hand, its $\gamma$--ray spectral index above 10 GeV ($\alpha_{\gamma}=1.88$) is too low for typical FSRQs, but rather normal for BL Lac objects \citep{Ackermann13}. 

In conclusion, although no other high--energy emitting source was found within the 1FHL positional uncertainty ellipse, we cannot rule out the possibility that this quasar is a background object and that the potential association is actually spurious. To conclusively pinpoint the true association it is necessary to obtain a spectrum of the optical counterpart of the above mentioned radio source, which shows a magnitude $R$ of $\sim\,19.5$ in the USNO-B1.0 catalogue.

\section{Conclusions}
We obtained optical spectra for 14 potential associations with $\gamma$--ray sources from the 1FHL catalogue, which were selected on the basis of their X--ray emission. These are our findings:

   \begin{enumerate} 
   \item  From these spectra, it is clear that 12 of these objects correspond to blazars belonging to the BL Lac class, with non-thermal continua and no spectral features. There are two exceptions: U0750-02519189, associated with 1FHL\,J0639.6-1244, whose host galaxy's spectroscopic signature is visible and allowed us to place it at a redshift of $z=0.135\pm 0.001$; and U1530-0317394, associated with 1FHL\,J2002.6+6303, which presents absorption from an intervening medium, placing it at a minimum redshift $z\ge 0.9$. The other 10 BL Lacs remain without a value for their redshifts.
   \item At least 8 out of the 12 BL Lacs present spectral indices in agreement with a synchrotron peak at a frequency higher than $10^{15}\rm{Hz}$, meaning they are likely candidates to be detected at TeV energies. 
   \item The X--ray counterpart within the field of 1FHL\,J1115.0-0701 presents strong, broad optical emission lines at a redshift of $z=2.929\pm 0.003$, indicating that it is an AGN of the quasar class. By measuring the flux and width of the CIV emission line, we could estimate the mass of the central black hole as $4.6\times10^{9}\,\rm{M}_{\sun}$. Assuming this is the real counterpart for 1FHL\,J1115.0-0701, its luminosity would be $L_{\gamma}=3.2\times10^{47}\,\rm{erg/s}$ and $L_X=7.9\times10^{45}\,\rm{erg/s}$. However, from multiwavelength considerations, we cannot rule out the possibility that this quasar is a background object and that its potential association with the $\gamma$--ray source is the product of statistical contamination. Further analysis is needed, in particular concerning the other object proposed as the real counterpart, radio source NVSS\,J111511--070238.
   \item U1575--03416943, potentially associated with 1FHL\,J1410.4+7408 A, shows relatively narrow but strong emission lines at a redshift of $z=0.429\pm 0.001$. Given its optical spectral characteristics, we classified it as a NLS1. For this object we infer a central black hole mass of $\sim5\times10^{6}\,\rm{M}_{\sun}$. 
 	\item Given that the source 1FHL\,J1410.4+7408 was potentially associated with objects A (a NLS1) and B (a BL Lac), we suggest -to be conservative- that it is still not clear which of the two sources is the actual $\gamma$--ray emitter, or if both of them are contributing to the total $\gamma$--ray emission. However, it is more likely the BL Lac object associated with 1FHL\,J1410.4+7408\,B.
   \item Our optical spectroscopy confirmed all the counterpart candidates of the X--ray sources potentially associated with 1FHL objects selected for this paper, with 1FHL\,J1115.0--0701 as the only possible exception. We were able to classify all of them as extragalactic high--energy active nuclei. This strengthens the utility of the proposed approach - crossmatching $\gamma$--ray positions to soft X--ray ones, improving accuracy, then completing the identification process with optical follow--up work and multiwavelength archival data.
   
   \end{enumerate}

\begin{acknowledgements} 

E. J. Marchesini would like to thank Francesco Massaro and Paola Grandi for the useful discussions on this work, and Gianluca Israel for coordinating the NOT observations and for useful comments. N. Masetti thanks the Facultad de Ciencias Astron\'omicas y Geof\'isicas de La Plata for the warm hospitality during the preparation of this paper. We thank Roberto Gualandi for night assistance at the Loiano telescope, and Gloria Andreuzzi for coordinating our service mode observation at the TNG.
We also acknowledge the Italian Space Agency financial support (ASI/INAF agreement n. 2013-025.R.0). This work is funded under the co-tutoring agreement between University of Turin and University of La Plata.
 \end{acknowledgements}

\bibliographystyle{aa}
\bibliography{Biblio}

\end{document}